\documentclass[12pt,a4paper]{article}

\title{Space-time models derived from Schwarzschild's solution\\
{\small }}

\author{Llu\'{\i}s\ Bel\\
\emph{wtpbedil@lg.ehu.es}
}
\date{}
\begin{document}

\maketitle

\begin{abstract}

We discuss two space-time models: one is expanding, the other is static. They are both derived from  Schwarzschild's exterior solution. But they differ in the implementation of the parallelism at a distance and the choice of their master frame of reference.

\end{abstract}

\section*{Introduction}

A space-time model is a three ingredients concept. The first one is a differential manifold ${\cal V}_4$, of class $C^\infty$ for example. The second is a 4-dimensional Riemannian metric of Lorentzian normal type:

\begin{eqnarray}
\label{2.1}
&& ds^2=g_{\alpha\beta}dx^\alpha dx^\beta = \eta_{ab}\theta^a_\alpha \theta^b_\beta \\
&& \hspace{-1cm}\alpha,\beta\cdots =0,1,2,3. \quad a,b,\cdots=0,1,2,3 \nonumber
\end{eqnarray}
where $\theta^a_\alpha$ are four linearly independent 1-forms, such that both $g_{\alpha\beta}$ and its inverse $g^{\alpha\beta}$ are of class $C^\infty$ on ${\cal V}_4$.

The third ingredient is a selection among all orthonormal decomposition  $\theta_a^\alpha$  of (\ref{2.1}) of a global one defined up to a constant Lorentz transformation. This serves two purposes: i) to define a parallelism at a distance allowing to compare two directions at different events and ii) to select a master global time-like vector field on which the meaning of the model partly resides. 

In Section 1 we consider Whitehead's form of Schwarzschild solution to be compared with more familiar writings of the corresponding line-elements. 

In section 2 we recall some recent results about the classification of Weizenb\"{o}ck's geometries associated with orthonormal decompositions, distinguishing in particular what we call the Doubly special one to be used in the sequel of the paper. 

In Section 3 we present an expanding space-time model, derived from a Doubly special orthonormal decomposition. And in Section 4 we discuss the familiar static model, on which rest the success of General relativity at the celestial mechanics level, including the consideration of an orthonormal decomposition subordinated to the Doubly special one of introduced in the preceding section.

This paper does not contain many new technical results. Our purpose was to suggest, using as an example Schwarzschild's exterior solution, the potential polymorphism lying behind every local solution of Einstein's field equations. 


\section{Gravitation described as a retarded interaction}

{\it Whitehead's line-element} Let us consider a point particle of mass $m$ moving with constant velocity in the framework of Special relativity. In 1922 Whitehead \cite{Whitehead}, wishing to propose a new theory of gravitation, considered the following tensor:

\begin{equation}
\label{1}
g_{\alpha\beta}=\eta_{\alpha\beta}+\frac{2m}{\hat r^3}\hat L_\alpha\hat L_\beta 
\end{equation}
as a description of the gravitational field of such particle. The definitions are as follows:

$u^\alpha$ is the unit time-like vector tangent to the world-line $\cal U$ of the particle.

$x^\alpha$ is the event where the field is calculated.

$\hat x^\alpha$ is the intersection of the past null cone, with vertex at $x^\alpha$, with the world-line $\cal U$. 

$\hat L^\alpha=x^\alpha-\hat x^\alpha$ is a null vector and $\hat r=-u_\alpha \hat L^\alpha$ is a positive definite distance from $x^\alpha$ to $\cal U$.

Keep in mind also that some calculations in this paper require the use of the following derivatives of $\hat L^\alpha$ and $\hat r$ taking into account that $\hat x^\alpha$ is a functional of $x^\alpha$:

\begin{equation}
\label{1.0}
\hat\partial_\alpha\hat L^\beta=\delta ^\beta_\alpha+\frac{1}{\hat r}\hat L^\beta \hat L_\alpha \quad
\hat\partial_\alpha\hat r=-u_\alpha +\frac{1}{\hat r}\hat L_\alpha
\end{equation}
 
{\it Droste´s line-element}\,\cite{Droste}.-
Whitehead dealt with (\ref{1}) as if these quantities were the coefficients of a Lorentzian metric:

\begin{equation}
\label{2}
ds^2=\left(\eta_{\alpha\beta}+\frac{2m}{\hat r^3}\hat L_\alpha\hat L_\beta\right) dx^\alpha dx^\beta
\end{equation}
but he failed to notice that his line-element, for a single point source, was covariantly equivalent to Schwarzschild's solution of Einstein's equations. This equivalence was pointed out by Eddington \cite{Eddington} in a very short Letter to the Editor of Nature that we recall below.
Let us choose a frame of reference such that the particle is at rest and $\hat x^i=0$. Then, dropping the retarded symbol $\hat{}$ :

\begin{equation}
\label{3}
u^0=1, \ u^i=0, \ L^0=r=|\vec x|, \ L^i=x^i \\ 
\end{equation}
and the line-element (\ref{2}) becomes:

\begin{equation}
\label{4}
ds^2=-\left(1-\frac{2m}{r}\right)dt^2-\frac{4m}{r^2}x_i dx^i dt+\left(\delta_{ij}+\frac{2m}{r^3}x_ix_j\right)dx^idx^j
\end{equation}

This line-element can be further transformed implementing two simplifications. If we first use polar coordinates we get the non diagonal metric:

\begin{equation}
\label{5}
ds^2=-\left(1-\frac{2m}{r}\right)dt^2-\frac{4m}{r} dr dt+\left(1+\frac{2m}{r}\right)dr^2+r^2d\Omega^2
\end{equation}
and finally, with an adapted time substitution\,\footnote{$f(x)\leftarrow x$ is a coordinate transformation $x=f(x^\prime)$ and setting $x^\prime=x$ in the result}:

\begin{equation}
\label{6}
t-2m\ln\left(\frac{2m}{r-2m}\right)\leftarrow t
\end{equation} 
we obtain:

\begin{equation}
\label{7}
ds^2=-\left(1-\frac{2m}{r}\right)dt^2+\left(1-\frac{2m}{r}\right)^{-1}dr^2+r^2d\Omega^2
\end{equation}
which is nowadays the most used form of Schwarzschild´s local solution.

Although Whitehead's theory is no match to Einstein's General relativity\,\footnote{See for instance a vindication with no nuances of this point of view in \cite{Gibbons}. Whitehead's point of view in the framework of the linear approximation of General relativity was discussed in \cite{Bel06}}, his line-element (\ref{2}) gives in several respects a better description of the gravitational field of a point-particle than that given by any other covariantly equivalent to it.

{\it Causal structure}: (\ref{2}) emphasizes the fact that gravitation is a causal interaction. This meaning that if at some moment the particle m departs from its constant velocity motion at an event $x_e^\alpha$ this will be felt only at those latter events $x_a^\alpha$ with $x_a^0=x_e^0+|\vec x_a-\vec x_e|$.

Moreover using (\ref{4}) or (\ref{5}) we benefit in General relativity of using coordinates with an unambiguous meaning, coming from the Special relativity model, to discuss for instance the causal structure of the space-time. Let us consider the radial null lines of the line-element (\ref{5}). Using the condition $ds^2=0$ we get the following two vectors, for each value of the distance $r$:

\begin{eqnarray}
\label{7.0.1}
l^0&=&1, \ l^1=1 \\
\label{7.0.2}
k^0&=&1+\frac{2m}{r}, \ k^1=-1+\frac{2m}{r}
\end{eqnarray}
where we have chosen an arbitrary factor so that they are future pointing vectors ($l^0>0,k^0>0$) and so that ($l^0=1,k^0=1$) for $r=\infty$.
Notice that the vector (\ref{7.0.1}) always points outwards $l^1>0$ and that the vector (\ref{7.0.2}) points inwards $k^1<0$ when $r>2m$ but points outwards $k^1>0$ when $r<2m$. Therefore any event at $r_0>0$ can be the cause of an event happening at $r_1>2m$. And if $r_0<2m$ then it can be twice the cause, so to speak, of another event with $r_0<r_1<2m$, through two signals propagating along two different null geodesics.

{\it Motion of the source}: For example if instead of assuming that the particle is at rest with respect to an observer with fixed location $x^i$  we assume that it is moving with velocity $v$ along, say, the $x^1$ axis, i.e, if instead of assuming the first two equations (\ref{3}) we assume that:

\begin{equation}
\label{7.1}
u^0=\frac{1}{\sqrt{1-v^2}}, \quad u^1=\frac{v}{\sqrt{1-v^2}}, \quad u^2=u^3=0
\end{equation}
then we have:

\begin{equation}
\label{7.2}
\hat L^0=-\hat L^1=\frac{r v}{\sqrt{1-v^2}}, \quad \hat L^2=\hat L^3=0
\end{equation}
where $r=x^1$, and we get at the end along the axis $x^1$ the following restriction of the line-element (\ref{4}).

\begin{equation}
\label{7.3}
ds^2=-\left(1-\frac{2m(v)}{r}\right)dt^2-\frac{4m(v)}{r}dr dt+\left(1+\frac{2m(v)}{r}\right)dr^2 
\end{equation}
with:

\begin{equation}
\label{7.4}
m(v)=m\frac{(1-v)^2}{(1-v^2)^{1/2}}
\end{equation}
revealing the dependence of the particle fiduciary mass on its velocity.

{\it Singularity}: The potential components (\ref{1}) as well as the components of the inverse tensor:

\begin{equation}
\label{7.5}
g^{\alpha\beta}=\eta^{\alpha\beta}-\frac{2m}{\hat r^3}\hat L^\alpha\hat L^\beta
\end{equation}
have a unique singularity located on the world-line of the particle which is the source of the gravitational field.

As we have mentioned Eddington recognized the covariant equivalence of the two line-elements (\ref{5}) and (\ref{7}) but he presented his finding as a way to derive (\ref{5}) from (\ref{7}). For this reason nowadays some relativists refer to Whitehead's line-element as Eddington-Finkelstein's extension of Schwarzschild's line-element. This is twice incorrect because Eddington did not discover (\ref{7}) and the line-element that Finkelstein invented \cite{Finkelstein} is the Whitehead one with the trivial substitution $t-r \leftarrow t$ that does not bring any improvement whatsoever. 

To end this section let us remind the reader that the line-element (\ref{7}) is also known as the Kerr-Schild \cite{Kerr} form of Schwarzschild's solution. Kerr-Schild considered also the general class of metrics of the following form:

\begin{equation}
\label{7.6}
g_{\alpha\beta}=\eta_{\alpha\beta}+ K_\alpha K_\beta, \
\end{equation} 
$K_\alpha$ being a null vector with respect to Minkowski's and Riemannian's metrics. Among many others interesting properties let us keep in mind that:

\begin{equation}
\label{C.1}
\det|g_{\alpha\beta}|=-1
\end{equation}


\section{Special Weitzenb\"{o}ck connections}

To each orthonormal decomposition of (\ref{2.1}) corresponds a Weitzenb\"{o}ck connection defined by:

\begin{equation}
\label{3.2}
\widetilde\Gamma^\lambda_{\beta\gamma}=e^\lambda_a\partial_\gamma\theta^a_\beta
\end{equation}
where $e^\alpha_b$ is the vector frame dual of $\theta^a_\alpha$:

\begin{equation}
\label{2.1.1}
\theta^a_\alpha e^\alpha_b=\delta^a_b
\end{equation}
Both $\theta^a_\alpha$ and $e_a^\alpha$ in (\ref{2.1.1}) have zero covariant derivatives:

\begin{equation}
\label{3.3.1}
\widetilde\nabla\theta^a_\alpha=0, \ \ \widetilde\nabla e^\alpha_b=0 
\end{equation}

The Riemann tensor of Weitzenb\"{o}ck's connections is zero:

\begin{equation}
\label{3.3.2}
\widetilde R^\alpha_{\beta\gamma\delta}=
\partial_\gamma\widetilde\Gamma^\alpha_{\beta\delta}-\partial_\delta\widetilde\Gamma^\alpha_{\beta\gamma}+ 
\widetilde\Gamma^\alpha_{\rho\gamma}\widetilde\Gamma^\rho_{\beta\delta}-\widetilde\Gamma^\alpha_{\rho\delta}\widetilde\Gamma^\rho_{\beta\gamma}=0
\end{equation} 
The Torsion and Contortion tensors are defined respectively by:

\begin{equation}
\label{3.4}
T^\lambda_{\beta\gamma}=-(\widetilde\Gamma^\lambda_{\beta\gamma}-\widetilde\Gamma^\lambda_{\gamma\beta}), \quad
K^\lambda_{\beta\gamma}=\widetilde\Gamma^\lambda_{\beta\gamma}-\Gamma^\lambda_{\beta\gamma}
\end{equation}
where $\Gamma^\lambda_{\beta\gamma}$ is the Christoffel connection of the Riemannian metric.

Let us consider the Fermi-Walker transport of a vector $P^\alpha$ along a time-like world-line $\cal W$ with tangent unit vector $w^\alpha$. This meaning that:

\begin{equation}
\label{6.20}
\frac{D P^\alpha}{d\tau}\equiv \frac{\nabla P^\alpha}{d\tau}+(a^\alpha w_\rho-w^\alpha a_\rho)P^\rho=0, \
a^\alpha=\frac{\nabla w^\alpha}{d\tau}                                      
\end{equation}
Using:

\begin{equation}
\label{6.21}
\frac{\nabla \theta^a_\alpha}{d\tau}=w^\sigma(\partial_\sigma\theta^a_\alpha-\Gamma^\rho_{\alpha\sigma} \theta^a_\rho),
\end{equation}
the definition of the Contortion tensor (\ref{3.4}), and (\ref{3.3.1}):

\begin{equation}
\label{6.21.1}
\widetilde\nabla_\sigma \theta^a_\alpha=\partial_\sigma\theta^a_\alpha-\widetilde\Gamma^\rho_{\alpha\sigma} \theta^a_\rho=0,
\end{equation}
we obtain:

\begin{equation}
\label{6.22}
\frac{\nabla\theta^a_\alpha}{d\tau}=K^\mu_{\alpha\sigma}\theta^a_\mu,
\end{equation}
Introducing now the physical scalar components:

\begin{equation}
\label{6.23}
P^a=P^\alpha\theta^a_\alpha, \ K^a_{bc}=K^\alpha_{\mu\nu}\theta^a_\alpha e^\mu_b e^\nu_c, \ \cdots
\end{equation}
the Fermi-Walker transport of $P^a$ becomes:

\begin{equation}
\label{6.24}
\frac{dP^a}{d\tau}+(a^a w_b-w^a a_b)P^b=K^a_{bc}w^cP^b
\end{equation}
telling us that the Contortion tensor is the torque of the precession of $P^a$ along the world-line $\cal W$.

Some Weitzenb\"{o}ck connections may define a parallelism at a distance if appropriate supplementary conditions are satisfied. In \cite{Bel08} we defined three particular classes of them:

We defined as Special ones those for which one has:

\begin{equation}
\label{3.4.1}
\eta_{ab}\theta^a\wedge d\theta^b=0
\end{equation}
A particular case being the integrable one:

\begin{equation}
\label{3.4.2}
\theta^a\wedge d\theta^a=0, \quad a=0,1,2,3
\end{equation}
For all these connections one has that:

\begin{equation}
\label{3.4.3}
T_{[\alpha\beta\gamma]}=0, \ T_{\alpha\beta\gamma}=g_{\gamma\rho}T_{\alpha\beta}^\rho
\end{equation}
which is equivalent to:

\begin{equation}
\label{3.4.4}
T_{\alpha\beta\gamma}=K_{\alpha\beta\gamma}, \ K_{\alpha\beta\gamma}=g_{\alpha\rho}K_{\beta\gamma}^\rho
\end{equation} 

Weitzenb\"{o}ck connections are intrinsic constructs depending on orthonormal decompositions unrelated to the coordinates being used, except in a particular case. We defined Doubly special Weitzenb\"{o}ck's connections as those connections that besides satisfying (\ref{3.4.3}) there exist an adapted coordinate system such that:

\begin{equation}
\label{A.0}
\widetilde\Gamma_{\beta\gamma\alpha}=\widetilde\Gamma_{\alpha\gamma\beta},\ \widetilde\Gamma_{\alpha\beta\gamma}=g_{\gamma\rho}\widetilde\Gamma_{\alpha\beta}^\rho 
\end{equation}
This is not meant to restrict the admissible system of coordinates. It means only that a special system of coordinates exist that it is convenient to use.
It is very easy to prove that if a Weitzenb\"{o}ck connection is integrable, which means that can be but in diagonal form,  then it is Doubly special.  

We proved in \cite{Bel08} that in the adapted system of coordinates one has:

\begin{equation}
\label{A.1}
\widetilde\Gamma^\alpha_{\beta\gamma}=\frac12 g^{\alpha\rho}\partial_\gamma g_{\beta\rho}, \ \ 
\widetilde\Gamma_{\beta\gamma\alpha}=\frac12\partial_\gamma g_{\beta\alpha}
\end{equation}
Consider the Christoffel symbols of a Riemannian metric (\ref{2.1}). They can be written as:

\begin{equation}
\label{A.1.1}
\Gamma^\alpha_{\beta\gamma}=\frac12 g^{\alpha\rho}\partial_\gamma g_{\beta\rho}-
\frac12 g^{\alpha\rho}(\partial_\rho g_{\beta\gamma}-\partial_\beta g_{\rho\gamma})
\end{equation}
Therefore, we see from (\ref{3.4}) that for these connections, in the adapted system of coordinates, we have:

\begin{equation}
\label{A.1.2}
K^\alpha_{\beta\gamma}=
\frac12 g^{\alpha\rho}(\partial_\rho g_{\beta\gamma}-\partial_\beta g_{\rho\gamma})
\end{equation} 
Since (\ref{A.1}) are in this case the symbols of a Weitzenb\"{o}ck connection the curvature tensor must be zero and a simple calculation shows that we must have:

\begin{equation}
\label{A.1.0}
\partial_\delta g^{\alpha\rho}\partial_\gamma g_{\beta\rho}
-\partial_\gamma g^{\alpha\rho}\partial_\delta g_{\beta\rho}=0
\end{equation}

Conversely we prove below that if a system of coordinates exists satisfying these equations then the symbols defined in (\ref{A.1}) are those of a Doubly special connection defined up to a global constant Lorentz transformation.

To obtain the corresponding orthonormal decompositions $\theta^a_\alpha$ of $g_{\alpha\beta}$ we proceed as follows: let us consider the following symbols\,\footnote{In 1914 Einstein considered these symbols to be the "components of the gravitational field", but dropped this interpretation latter on}:

\begin{equation}
\label{A.2}
C^\alpha_{\beta\gamma}=\frac12 g^{\alpha\rho}\partial_\gamma g_{\beta\rho}, \ \ C_{\beta\gamma\alpha}
=\frac12\partial_\gamma g_{\beta\alpha}
\end{equation}
Using (\ref{A.1.0}) a simple calculation shows that these symbols satisfy the following equations:

\begin{equation}
\label{A.3}
\partial_\gamma C^\alpha_{\beta\delta}-\partial_\delta C^\alpha_{\beta\gamma}+ 
C^\alpha_{\rho\gamma} C^\rho_{\beta\delta}- C^\alpha_{\rho\delta} C^\rho_{\beta\gamma}=0
\end{equation}

Let us consider the system of differential equations:

\begin{equation}
\label{A.5}
\partial_\gamma\theta^b_\beta=\theta^b_\rho C^\rho_{\beta\gamma}
\end{equation}
From (\ref{A.3}) it follows that this system is completely integrable. Therefore, a particular event $x_0$ being selected, we can always choose the solution defined by initial conditions $\theta^a_\alpha(x_0)$
such that:

\begin{equation}
\label{A.6}
g_{\alpha\beta}(x_0)= \eta_{ab}\theta^a_\alpha(x_0)\theta^b_\beta(x_0).
\end{equation}

Let us now define, using the solution thus selected, the quantities:

\begin{equation}
\label{A.7}
s_{\alpha\beta}= \eta_{ab}\theta^a_\alpha\theta^b_\beta.
\end{equation}
Differentiating these equations and using (\ref{A.5}) we obtain:

\begin{equation}
\label{A.8}
\partial_\gamma s_{\alpha\beta}=s_{\rho\beta}C^\rho_{\alpha\gamma}+s_{\rho\alpha}C^\rho_{\beta\gamma}
\end{equation}
Again, from (\ref{A.3}), it follows that this system of equations is completely integrable and choosing initial conditions such that:

\begin{equation}
\label{A.9}
s_{\alpha\beta}(x_0)=g_{\alpha\beta}(x_0)
\end{equation}
we get a unique solution that coincides with the tensor components we started with:

\begin{equation}
\label{A.10}
s_{\alpha\beta}=g_{\alpha\beta}
\end{equation}
This is so because $g_{\alpha\beta}$ obviously satisfies (\ref{A.9}) by construction and because a short calculation using (\ref{A.2}) shows that it is a solution of (\ref{A.8}).

Multiplying now (\ref{A.5}) by the elements of the dual basis $\theta^b_\alpha$, we obtain the Weitzenb\"{o}ck's connection symbols:

\begin{equation}
\label{A.11}
\widetilde\Gamma^\alpha_{\beta\gamma}\equiv e^\alpha_a\partial_\gamma\theta^a_\beta=C^\alpha_{\beta\gamma}, \ \
\widetilde\Gamma_{\beta\gamma\alpha}=C_{\beta\gamma\alpha}
\end{equation}
from where, using (\ref{A.2}), it follows that the symmetry property (\ref{A.0}) is satisfied.

Therefore to every system of coordinates satisfying (\ref{A.1.0}) we can always associate to it a particular Special Weitzenb\"{o}ck connection which is a Doubly special one. 

Let us consider the line-element (\ref{1}). A first calculation proves that the conditions (\ref{A.1.0}) are satisfied. A second calculation proves that the corresponding Weitzenb\"{o}ck connection is that derived from the following orthonormal decomposition:

\begin{equation}
\label{3.5}
\theta^a_\alpha=\delta^a_\alpha+\frac{m}{\hat r^3}\hat L^a\hat L_\alpha,  \quad e^a_\alpha=\delta^a_\alpha-\frac{m}{\hat r^3}\hat L_a\hat L^\alpha
\end{equation}
Both results holds may be easily generalized for all those metrics considered in (\ref{7.6}).  


\section{The expanding model}

One of the goals of this paper is to show that the same local solution of Einstein's field equations may lead to several distinct global space-time models. We shall discuss in the sequel two particular ones derived from the Schwarzschild vacuum solution. Both share the same abstract manifold $R\times(R^3-{x^i_0})$ but use different orthonormal decompositions of the metric to implement the space-time parallelism at a distance, and also different master frames of reference.

The first of these models that we consider is derived from Whitehead's line element (\ref{2}) of the Schwarzschild solution  and the orthonormal decomposition (\ref{3.5}), corresponding to a Doubly special Weitzenb\"{o}ck connection. Accordingly the master frame of reference will be defined as the global global time-like vector field $e_0^\alpha$.

In a frame of reference of Minkowski space-time model such that we have (\ref{3})
the contravariant components of $e_0^\alpha$ in polar coordinates are:

\begin{equation}
\label{4.1}
e_0^0=1+\frac{m}{\hat r}, \quad e_0^1=\frac{m}{\hat r}
\end{equation}
The trajectories of $e_0^\alpha$ are therefore the solutions of the differential equation:

\begin{equation}
\label{4.2}
\frac{dt}{dr}=-\frac{m}{r+m}
\end{equation}
whose general solution is:

\begin{equation}
\label{4.3}
t-t_0=\frac{1}{2m}(r+r_0+2m)(r-r_0)
\end{equation}
All these trajectories start at $r=0$ and end at $r=\infty$

The Christoffel connection in the native system of coordinates is:

\begin{equation}
\label{5.1}
\Gamma^\alpha_{\beta\gamma}=-\frac{m}{\hat r^4}u^\alpha \hat L_\beta \hat L_\gamma+\frac{m}{\hat r^3}\left(2\eta_{\beta\gamma}+\frac{3}{\hat r}(u_\beta \hat L_\gamma+\hat L_\beta u_\gamma)\right)\hat L^\alpha -\frac{m}{\hat  r^5}\left(3+\frac{2m}{\hat r}\right) \hat L^\alpha \hat L_\beta \hat L_\gamma
\end{equation}
from where we derive in particular:

\begin{equation}
\label{5.2}
\Gamma^\alpha_{\beta\alpha}=0
\end{equation}
that follows also from (\ref{C.1}). The Weitzenb\"{o}ck connection is:

\begin{equation}
\label{5.3}
\widetilde\Gamma^\alpha_{\beta\gamma}=
\frac{m}{\hat r^3}(\delta^\alpha_\gamma \hat L_\beta+\eta_\beta\gamma\hat L^\alpha)+\frac{m}{\hat r^4}u^\alpha \hat L_\beta \hat L_\gamma+
\frac{m}{\hat r^4} \hat L^\alpha\left(u_\beta \hat L_\gamma+3u_\gamma\hat L_\beta-\frac{3}{\hat r}\hat L_\beta \hat L_\gamma\right)
\end{equation}
from where we derive the Torsion tensor:

\begin{equation}
\label{5.4}
T^\alpha_{\beta\gamma}=\frac{m}{\hat r^3}(\delta^\alpha_\beta\hat L_\gamma-\delta^\alpha_\gamma \hat  L_\beta)+\frac{2m}{\hat r^4}\hat L^\alpha
(u_\beta \hat L_\gamma-u_\gamma \hat L_\beta)
\end{equation}

Using (\ref{3.5}) and (\ref{5.2}) we easily obtain:

\begin{equation}
\label{5.5}
\nabla_\alpha e_0^\alpha=-\frac{m}{\hat r^3}\hat L_0
\end{equation}
Since $L_0=-(x^0-\hat x^0)$ is negative this means that the main time-like congruence is expanding at a decreasing rate when $\hat r$ increases so that the model remains asymptotically Minkowskian at infinity.

{\it Adapted coordinates} The Riemannian metric of the quotient space, where each point corresponds to a trajectory of the vector field $e_0^\alpha$ is :

\begin{equation}
\label{5.6}
d\hat s^2=\delta_{ij}\theta^i\theta^j,   \ \theta^i=\theta^i_\alpha dx^\alpha
\end{equation} 
To make explicit the three dimensions we need to find adapted coordinates such that three of them remain constant along every single trajectory of $e_0^\alpha$. To this end we proceed to make a coordinate transformation of the following type:

\begin{equation}
\label{5.7}
t^\prime =t, \ x^{\prime i}=x^i R(t,r)\Rightarrow r^\prime=R(t,r)r
\end{equation} 
such that:

\begin{equation}
\label{5.8}
e^{\prime j}_0=\left(1+\frac{m}{r}\right)\frac{\partial x^{\prime j}}{\partial t}+\frac{m}{r^2}x^i\frac{\partial x^{\prime j}}{\partial x^i}=0
\end{equation}
Or equivalently:

\begin{equation}
\label{5.9}
\left(1+\frac{m}{r}\right)\frac{\partial R}{\partial t}+\frac{m}{r}\frac{\partial R}{\partial r}+\frac{m}{r^2}R=0
\end{equation}

Let us consider a particular event ($t_0,x^i_0)$. The general solution of the preceding equation such that $t^\prime_0=t_0,\ x^{\prime i}_0=x^i_0$ is:

\begin{equation}
\label{5.10}
R=\frac{r_0}{r}\exp(\mu(t-t_0)-\frac{\mu}{2m}(r^2-r_0^2)-\mu(r-r_0))
\end{equation}
where $\mu$  is an arbitrary constant that we shall fix later. The relationship between  $r$ and $r^\prime$ is:

\begin{equation}
\label{5.11}
r=-m+\left((m+r_0)^2+\frac{2m}{\mu}(\ln(r_0)-\ln(r^\prime)+\mu(t-t_0))\right)^{1/2}
\end{equation}

The line-element (\ref{5}) becomes then:

\begin{equation}
\label{5.12}
ds^2=-\frac{r^2}{(r+m)^2}{dt^\prime}^2+\frac{m^2(r+2m)}{\mu^2r{r^\prime}^2(r+m)^2}{dr^\prime}^2
+\frac{2m^2}{\mu r^\prime(r+m)^2}dt^\prime dr^\prime +r^2d\Omega^2
\end{equation}
And the corresponding quotient metric is:

\begin{equation}
\label{5.13}
d\hat s^2=\frac{m^2}{\mu^2r^2{r^\prime}^2}dr^{\prime 2}+r^{\prime 2}d\Omega^2
\end{equation}
Choosing the coefficient of $dr^2$ to be $1$ at the chosen event of reference gives:

\begin{equation}
\label{5.14}
\mu=\frac{m}{rr^\prime}
\end{equation}

We want now to explore some of the physics of the model that could be used in principle to disprove the model.

{\it The strength of gravity}. The intrinsic curvature $b_\alpha$ of the world-lines of the congruence $e_0^\alpha$ can be calculated using the formula:

\begin{equation}
\label{5.15}
f=i(e_0)d\theta^0
\end{equation}
We obtain thus:

\begin{equation}
\label{5.16}
f_\alpha=-\frac{m}{r^3}\left(1+\frac{m}{r}\right)L_\alpha+\frac{m}{r^3}(\delta^0_\alpha+2u_\alpha)
\end{equation}
From where we derive the scalar force per unit mass:

\begin{equation}
\label{5.17}
f=\sqrt{g^{\alpha\beta}f_\alpha f_\beta}=\frac{m}{r^2}
\end{equation}

{\it Red-shifts}. Let us consider a radial light-ray of frequency $\nu_0$ originated at an event ($r_0, t_0$) and reaching the location $r_1>r_0$ with $r_1>2m$ at time $t_1>t_0$. The tangent vector to this null geodesic is the global vector $l^\alpha$ defined in (\ref{7.0.1}). And the reception frequency $\nu_1$ can be calculated as usual using the variation equation:

\begin{equation}
\label{5.18}
\delta t_1-\delta t_0=\delta r_1-\delta r_0
\end{equation}
as well as the variation equations derived from the differential equations (\ref{4.2}):

\begin{equation}
\label{5.19}
\delta r=\frac{m}{r+m}\delta t
\end{equation}
and the relationship between the coordinate time $t$ and proper-time along the world-line $e_0^\alpha$:

\begin{equation}
\label{5.20}
\delta \tau=\frac{r}{r+m}\delta t
\end{equation} 
The red-shift obtained using $d\tau_1/d\tau_0=\nu_0/\nu_1$ is:

\begin{equation}
\label{5.21}
z=\frac{\nu_0}{\nu_1}-1=0.
\end{equation} 

If we assume now that $r_0<r_1$ with $r_0>2m$ then the null geodesics that we have to consider is the incomplete one whose tangent vector is the vector $k^\alpha$ defined in (\ref{7.0.2}). In this case the calculation of the red-shift follows the same lines as above replacing (\ref{5.18}) by:

\begin{equation}
\label{5.22}
\delta t_1-\delta t_0=-\frac{r_1+2m}{r_1-2m}\delta r_1+\frac{r_0+2m}{r_0-2m}\delta r_0
\end{equation}
The corresponding result is now:

\begin{equation}
\label{5.23}
z=\frac{\nu_0}{\nu_1}-1=\frac{r_0(r_1-2m)}{r_1(r_0-2m)}
\end{equation} 
This is also the red-shift formula when $2m>r_1>r_0$.
   

\section{The static model}

Most derivations of covariantly equivalent line-elements of Schwarzschild's vacuum solution assume from the beginning its time independence. Whitehead's formulation instead assumes that the four-velocity $u^\alpha$ of the source is constant in Minkowski's space-time.The equivalence of these two points of view follows from the fact that if we consider $u^\alpha$ as a constant vector field  defined at each event of ${\cal V}_4=R\times(R^3-\{r>0\})$ then a short calculation proves that:

\begin{equation}
\label{6.1}
Lie(u)g_{\alpha\beta}= u^\rho\partial_\rho g_{\alpha\beta}=0 
\end{equation}
which means that $u^\alpha$ is a Killing symmetry of the tensor field (\ref{1}). Using coordinates such that (\ref{3}) are satisfied we have:

\begin{equation}
\label{6.2}
g_{\alpha\beta}u^\alpha u^\beta=-1+\frac{2m}{\hat r}
\end{equation}
Therefore, in the sense of the Riemannian metric $g_{\alpha\beta}$,   $u^\alpha$ is space-like if $\hat r<2m$, is null if $\hat r=2m$ and it is time-like if $\hat r>2m$. In the sequel of this section we shall use the following notations:

\begin{equation}
\label{6.2.0}
\xi^\alpha(x)=u^\alpha, \ \xi=\sqrt{-g_{\alpha\beta}\xi^\alpha\xi^\beta}
\end{equation}

The usual static Schwarzschild's space-time model is a differential sub-manifold $\bar {\cal V}_4=R\times(R^3-B_3\{r>2m\})$ of ${\cal V}_4$ on which it is defined the restriction of the line-element (\ref{2}), or any other covariantly equivalent to it, as well as  a master frame of reference defined by the unit time-like vector field $\xi^{-1}\xi^\alpha$.

The first line-element that we shall use is Droste's one derived from (\ref{5}) by the substitution (\ref{6}) which is a global admissible time coordinate transformation on the manifold $\bar {\cal V}_4$. It can trivially be decomposed as:

\begin{equation}
\label{6.2.1}
ds^2= -(\theta^0)^2+d\sigma^2,  \ a,b,\cdots=1,2,3
\end{equation}
with:

\begin{equation}
\label{6.2.1.a}
\theta^0=\xi dt,  \ d\sigma^2=\frac{rdr^2}{r-2m}+r^2d\Omega^2
\end{equation}
where $d\sigma^2$ is the quotient metric defined on the manifold $\bar V_3=R^3-B_3\{r>2m\}$. Remember that $r$ was initially the retarded distance from the field event to the point source, but we see from the  $d\sigma^2$ above that the radial distance between two space points $r_0$ and $r_1$ is not $r_1-r_0$, and that when $r_0$ tends to $2m$ the distance from $r_1$ to $r_0$ tends to infinity. We believe that this is a new reason to add to those mentioned before\,\footnote{see, for example \cite{Bel07} and references therein} indicating that a metric conformal to $d\sigma^2$, namely:

\begin{equation}
\label{6.8}
d\bar\sigma^2=\xi^2d\sigma^2=dr^2+r(r-2m)d\Omega^2
\end{equation}
is more appropriate to describe the quotient space geometry. Since with this modification of the radial distance from a point with radial coordinate $r$ to a point with radial coordinate $2m$ becomes finite and equal to $r-2m$  nothing restricts the possibility of considering the substitution $r-2m \leftarrow r$. This brings the line-element (\ref{7}) to Brillouin's form \cite{Brillouin} :

\begin{equation}
\label{6.9}
ds^2=-\frac{r}{r+2m}dt^2+\frac{r+2m}{r}dr^2+(r+2m)rd\Omega^2
\end{equation} 
which exhibits a single singularity at r=0. Notice also that the preceding substitution is a diffeomorphism of ${\cal V}_4$ into $\bar {\cal V}_4$.

Equivalently to the substitution $r+2m\leftarrow r$ we can use the corresponding Cartesian one:

\begin{equation}
\label{6.9.a}
x^i\left(1+\frac{2m}{r}\right)\leftarrow x^i
\end{equation}
so that the line-element $ds^2$ becomes:

\begin{equation}
\label{6.14}
ds^2=-\xi^2dt^2+ \frac{1}{\xi^4}\left(\delta_{il}-\frac{2m\xi^2}{r^3}\right)x_i x_ldx^idx^l, \ \xi^2=\frac{r}{r+2m}
\end{equation}
from where we get the contravariant components:

\begin{equation}
\label{6.15}
g^{00}=-\frac{1}{\xi^2}, \ \ g^{jk}=\xi^4\left(\delta^{jk}+\frac{2m}{r^3}\right)x^j x^k,
\end{equation}

We complete this model including in it an orthonormal decomposition of (\ref{6.14}) subordinated to (\ref{3.5}). By this we mean an orthonormal decomposition that is the image of (\ref{3.5}) by the event dependent special Lorentz transformation that takes the vector$e^\alpha_0$ into $\xi^{-1}\xi^\alpha$. A straightforward calculation leads to the 1-forms:

\begin{equation}
\label{6.12}
\theta^0_0=\xi, \ \theta^s_j=\frac{1}{\xi^2}\delta^s_j+\left(1-\frac{1}{\xi}\right)\frac{x^s x_j}{r^2\xi},
\end{equation}
and corresponding vector frame:
 
\begin{equation}
\label{6.13}
e^0_0=\frac{1}{\xi}, \ e^i_s=\xi^2\delta^i_s+\xi(1-\xi)\frac{x^i x_s}{r^2},
\end{equation}

Using the definition (\ref{3.2}) we get the non zero components of the Weitzenb\"{o}ck connection:

\begin{equation}
\label{6.16}
\widetilde\Gamma^0_{0k}=\frac{mx_k}{r^3\xi^2}, \ \widetilde\Gamma^i_{jk}=\frac{\xi-1}{r^2}\delta^i_kx_j-
\frac{2m}{\xi^2r^3}\delta^i_jx_k+\frac{\xi-1}{r^2\xi}\delta_{jk}x^i+\frac{m(1+2\xi)}{r^5\xi}x^i x_j x_k
\end{equation}
from where, using the definition of Torsion, we get the non zero strict components:  

\begin{equation}
\label{6.17}
T^0_{0k}=-\frac{mx_k}{r^3\xi^2}, \ T^i_{jk}=C(\delta^i_kx_j-\delta^i_jx_k),\ C=1-\xi-\frac{2m}{r\xi^2}
\end{equation}
Their full covariant form are:

\begin{equation}
\label{6.18}
T_{0k0}=\frac{mx_k}{r^3}, \ T_{jki}=\frac{C}{r^2\xi^4}(\delta_{ik}x_j-\delta_{ij}x_k)
\end{equation}
It is then very easy to check that the conditions (\ref{3.4.3}) are satisfied, this meaning that we are dealing with a Special Weitzenb\"{o}ck connection. At the same time we can use  (\ref{3.4.4}) and we obtain thus the components of the Contortion:

\begin{equation}
\label{6.19}
K^0_{0k}=-\frac{mx_k}{r^3\xi^2}, \ K^j_{ki}=C{\xi^2}(\xi^2g_{ik}x^j-\delta^j_ix_k)
\end{equation}
When necessary the Christoffel symbols can be calculated using the definition of the Contortion given in (\ref{3.4}).


\section*{Conclusion}

We conclude  from our considerations that Schwarzschild's solution can be associated with two natural, but radically distinct, space-time models. One of them uses Whitehead's line-element, or anyone other globally covariantly equivalent, and its Master frame of reference is the time-like vector field of an orthonormal decomposition leading to a Doubly special Weitzenb\"{o}ck connection. The space of this model is expanding in such way that the red-shift formulas for radial light-rays could be useful to disprove the model.

The other model is static and uses the Brillouin line-element, or anyone other globally covariantly equivalent on the manifold ${\cal V}_4$, and an orthonormal decomposition whose time-like vector field is the generator of a global isometry. Up to now no one hundred percent reliable test has disproved this model, and to our knowledge it is the only one that can be thought of as the limit of a sequence of bounded spherical bodies with constant density when the radius shrinks to zero \cite{Aguirre}.


\end{document}